\documentclass[showpacs,aps,pra,preprint]{revtex4-1}

\usepackage{epsfig,amsmath}
\usepackage{subfigure}
\usepackage{graphicx}
\usepackage{dcolumn}
\usepackage{stmaryrd}
\usepackage{mathrsfs}
\usepackage{pifont}
\usepackage{amsthm}
\usepackage{amssymb}
\usepackage{bm}
\usepackage{latexsym}
\usepackage{hyperref}
\usepackage{color}

\begin{document}

\title{Quantum algorithm for obtaining the eigenstates of a physical system}
\author{Hefeng Wang}
\thanks{Correspondence to wanghf@mail.xjtu.edu.cn}
\affiliation{Department of Applied Physics, Xi'an Jiaotong University, Xi'an
710049, China}

\begin{abstract}
We propose a quantum algorithm for solving the following problem: given the
Hamiltonian of a physical system and one of its eigenvalues, how to obtain
the corresponding eigenstate? The algorithm is based on the resonance
phenomena. For a probe qubit coupled to a quantum system, the system
exhibits a resonance dynamics when the frequency of the probe qubit matches
a transition frequency in the system. Therefore the system can be guided to
evolve to the eigenstate with known eigenvalue by inducing resonance between
the probe qubit and a designed transition in the system. This algorithm can
also be used to obtain the energy spectrum of a physical system and can
achieve even a quadratic speedup over the phase estimation algorithm.
\end{abstract}

\maketitle

\emph{Introduction.}--Obtaining the eigenstates and energy spectrum of a
physical system is of fundamental importance in quantum physics and quantum
chemistry. In principle, the task can be achieved by solving the Schr\"{o}%
dinger equation of the system. In most cases, however, the Schr\"{o}dinger
equations can not be solved exactly, and numerical approaches such as full
diagonalization or Monte Carlo methods are not efficient in terms of the
size of the system on a classical computer. The quantum phase estimation
algorithm~(PEA)~\cite{kitaev} has been proposed for solving the following
problem efficiently: given an unitary operator $U$, and one of its
eigenstate $|\Psi \rangle $, how to estimate the phase factor $\theta $ of
the corresponding eigenvalue $e^{i\theta }$ of $U$? And later the PEA is
applied for solving the Schr\"{o}dinger equation of a system on a quantum
compute to obtain the energy eigenvalues and eigenstates of a physical
systemr~\cite{abrams, tra}. The success probability of the PEA is
proportional to the square of the overlap of the guess state with the real
eigenstate of the system. Adiabatic quantum evolution~(AQE) is another
quantum algorithm for preparing an eigenstate of a system~\cite{farhi}. In
AQE, however, one can only prepare the ground state of the system, and the
scaling the runtime of the algorithm remains an open question in the case
where the ground state of the system is degenerate.

In this paper, we propose a different quantum algorithm for obtaining an
arbitrary eigenstate of a physical system by asking the following question:
given the Hamiltonian of a system and one of its eigenvalues, how to obtain
the corresponding eigenstate of the system? This algorithm is based on the
resonance phenomena that for a probe qubit coupled to a physical system, the
probe exhibits a dynamical response when it resonates with a transition in
the system. Therefore the system can be guided to evolve to the eigenstate
with known eigenvalue by inducing a resonance between the probe qubit and a
transition in the system. The algorithm can even achieve a quadratic speedup
over the PEA, and can also be used to obtain the energy spectrum of a system.

\emph{The algorithm.}--Without loss of generality, we illustrate the
algorithm by showing how to obtain the ground state of a physical system
provided the ground state energy is already known. Details of the algorithm
are as follows.

We construct a quantum register $R$ of $\left( n+1\right) $ qubits, which
contains one ancilla qubit and an $n$-qubit quantum register that represents
a physical system of dimension $N=2^{n}$. A probe qubit is coupled to $R$
and the Hamiltonian of the entire $\left( n+2\right) $-qubit system is in
the form
\begin{equation}
H=-\frac{1}{2}\omega \sigma _{z}+I_{2}\otimes H_{R}+c\sigma _{x}\otimes B,
\end{equation}%
where $I_{2}$ is the two-dimensional identity operator, $\sigma _{x}$ and $%
\sigma _{z}$ are the Pauli matrices. The first term in the above equation is
the Hamiltonian of the probe qubit, the second term is the Hamiltonian of
the register $R$, and the third term describes the interaction between the
probe qubit and $R$. Here, $\omega $ is the frequency of the probe qubit~($%
\hbar =1$), and $c$ is the coupling strength between the probe qubit and $R$%
, and $c\ll \omega $. The Hamiltonian of $R$ is in the form
\begin{equation}
H_{R}=|0\rangle \langle 0|\otimes \left[ \varepsilon _{0}\left( |0\rangle
\langle 0|\right) ^{\otimes n}\right] +|1\rangle \langle 1|\otimes H_{S},
\end{equation}%
where $H_{S}$ is the Hamiltonian of the system and $\varepsilon _{0}$ is a
parameter that is set as a reference point to the ground state energy $E_{1}$
of $H_{S}$. $B$ is an operator that acts on the register $R$, which can be
varied for different systems. The operator $B=\sigma _{x}\otimes A$, and $A$
acts on the state space of the system. The construction of operator $A$
depends on the system and will be discussed in the following sections.

To run the algorithm, first we prepare the probe qubit in its excited state $%
|1\rangle $ and the register $R$ in a reference state $|\Phi \rangle
=|0\rangle ^{\otimes \left( n+1\right) }$, which is an eigenstate of $H_{R}$
with eigenvalue $\varepsilon _{0}$, the $(n+2)$ qubits are in state $|\Psi
_{0}\rangle =|1\rangle |\Phi \rangle =|1\rangle |0\rangle |0\rangle
^{\otimes n}$. Then evolve the entire $(n+2)$-qubit system with the
Hamiltonian $H$ for time $t$. After that, perform a measurement on the probe
qubit in basis of $|0\rangle $. When the probe qubit decays to its ground
state $|0\rangle $, the last $n$ qubits of the register $R$ evolves to the
ground state of the system with large probability. The circuit for the
algorithm is shown in Fig.~$1$.

In basis of \{$|\Psi _{0}\rangle =|1\rangle |0\rangle |0\rangle ^{\otimes n}$%
, $|\Psi _{i}\rangle =|0\rangle |1\rangle |\varphi _{i}\rangle $, $%
i=1,\cdots ,N$\}, where $|\varphi _{i}\rangle $ are the eigenstates of $%
H_{S} $ with the corresponding eigenvalues $E_{i}$, the Hamiltonian $H$ in
Eq.~($1$) is in the form: $H_{00}=\frac{1}{2}\omega +\varepsilon _{0}$; $%
H_{0i}=H_{i0}=c\langle \varphi _{i}|A|0\rangle ^{\otimes n}$, and $H_{ii}=-%
\frac{1}{2}\omega +E_{i}$, for $i\geq 1$; and $H_{ij}=0$ for $i,j\geq 1$ and
$i\neq j$. The ground state $|\varphi _{1}\rangle $ of $H_{S}$ is encoded in
the basis state $|\Psi _{1}\rangle =|0\rangle |1\rangle |\varphi _{1}\rangle
$. With the initial state being set as $|\Psi _{0}\rangle $, the Schr\"{o}%
dinger equation $i\frac{d}{dt}|\Psi \rangle =H|\Psi \rangle $ describes the
evolution of the entire $(n+2)$-qubit system from $|\Psi _{0}\rangle $ to
states $|\Psi _{i}\rangle =|0\rangle |1\rangle |\varphi _{i}\rangle $
through $N$ independent channels.

When the parameter $\varepsilon _{0}$ satisfies the condition $%
E_{1}-\varepsilon _{0}=\omega $, which means the transition frequency
between the reference state and the state $|\Psi _{1}\rangle $ matches the
frequency of the probe qubit, we have $H_{00}=H_{11}=\frac{1}{2}\omega
+\varepsilon _{0}$, and the system evolves from the initial state $|\Psi
_{0}\rangle $\ to the state $|\Psi _{1}\rangle =|0\rangle |1\rangle |\varphi
_{1}\rangle $ reaches maximal probability at time $t\sim 1/\left( c|\langle
\varphi _{1}|A|0\rangle ^{\otimes n}|\right) $, provides that the energy gap
between the ground state and the first excited state of the system is
finite. Then the last $n$ qubits of the register $R$ evolves to the ground
state $|\varphi _{1}\rangle $ of the system with high probability. The
evolution time $t$ is the runtime of the algorithm, and will be discussed in
the next section.

\emph{Efficiency of the algorithm.}--The efficiency of the algorithm depends
on the runtime $t$ and the probability of the system being evolved to state $%
|\Psi _{1}\rangle $ which encodes the ground state of the physical system, $%
P=|\langle \Psi _{1}|U(t)|\Psi _{0}\rangle |^{2}$. In general, we can not
solve the Schr\"{o}dinger equation exactly to obtain $P(t)$ of the
algorithm, but we can estimate the runtime $t$ by considering some special
cases.

In the algorithm, when the frequency of the probe qubit matches the
transition frequency between the reference state $|\Phi \rangle $ and the
eigenstate $|1\rangle |\varphi _{1}\rangle $ of $H_{R}$, the probability of
the $(n+2)$-qubit system being transferred from the initial state $|\Psi
_{0}\rangle $ to the state $|\Psi _{1}\rangle $ reaches maximum at certain
time $t$. There is also a probability for the system being transferred to
other states $|\Psi _{j}\rangle $, $j\!=2,\ldots ,N$. By applying the
first-order perturbation theory, this probability can be formulated as~\cite%
{cohn}
\begin{equation}
\sin ^{2}\!\left( \!\frac{\Omega _{0j}\tau }{2}\!\right) \!\!\frac{Q_{0j}^{2}%
}{Q_{0j}^{2}\!+\!\left( \!E_{j}\!-\!\varepsilon _{0}\!-\!\omega \!\right)
^{2}},\text{ }j\!=\!2,\ldots ,N
\end{equation}%
where $Q_{0j}=2c|\langle \varphi _{j}|A|0\rangle ^{\otimes n}|$, and $\Omega
_{0j}=\sqrt{Q_{0j}^{2}+\left( E_{j}-\varepsilon _{0}-\omega \right) ^{2}}$.
From the above equation one can see that as the transition frequency between
the reference state and the state $|1\rangle |\varphi _{j}\rangle $ becomes
closer to the frequency of the probe qubit, the probability of the system
being evolved to state $|\varphi _{j}\rangle $ is higher. Based on this
analysis, the runtime of the algorithm must be in between of the two assumed
special cases of the system: all the excited states $|\varphi _{j}\rangle $,
($j\!=2,\ldots ,N$) are degenerate at the lowest or the highest possible
energy levels of the system. By assuming that the ground state of the system
is non-degenerate and the excited states are $\left( N-1\right) $-fold
degenerate, we can calculate $P(t)$ by exactly solving the Schr\"{o}dinger
equation.

In the algorithm, the state $A|0\rangle ^{\otimes n}$ can be expanded by the
complete set of the eigenstates of the system $\{|\varphi _{i}\rangle
,i=1,2,\cdots ,N\}$ as $A|0\rangle ^{\otimes
n}=\sum\nolimits_{i=1}^{N}d_{i}|\varphi _{i}\rangle $, where $d_{i}=\langle
\varphi _{i}|A|0\rangle ^{\otimes n}$ and $\sum%
\nolimits_{i=1}^{N}|d_{i}|^{2}=1$. Suppose the excited states of the system
are $\left( N-1\right) $-fold degenerate with eigenvalue $E^{\prime }+\frac{1%
}{2}$, let $|\Psi _{2}\rangle =|0\rangle |1\rangle \frac{1}{\sqrt{N-1}}%
\sum_{i=2}^{N}|\varphi _{i}\rangle $ and $d_{1}=d$, the Hamiltonian matrix
of $H$ in basis $\{|\Psi _{0}\rangle =|1\rangle |0\rangle |0\rangle
^{\otimes n},|\Psi _{1}\rangle =|0\rangle |1\rangle |\varphi _{1}\rangle
,|\Psi _{2}\rangle =|0\rangle |1\rangle \frac{1}{\sqrt{N-1}}%
\sum_{i=2}^{N}|\varphi _{i}\rangle \}$ can be written as
\begin{equation}
H=\left(
\begin{array}{ccc}
\frac{1}{2}\omega +\varepsilon _{0} & cd & c\sqrt{1-|d|^{2}} \\
cd & \frac{1}{2}\omega +\varepsilon _{0} & 0 \\
c\sqrt{1-|d|^{2}} & 0 & E^{\prime }%
\end{array}%
\right) .
\end{equation}%
Let $|\Psi \left( t\right) \rangle =c_{0}\left( t\right) |\Psi _{0}\rangle
+c_{1}\left( t\right) |\Psi _{1}\rangle +c_{2}\left( t\right) |\Psi
_{2}\rangle $, the Schr\"{o}dinger equation with the above Hamiltonian can
be solved exactly and%
\begin{equation}
c_{1}\left( t\right) =4cd\sum_{x}\frac{(iE^{\prime }+x)e^{xt}}{%
12ix^{2}-8\left( E^{\prime }+1\right) x+4ic^{2}-4iE^{\prime }-i},
\end{equation}%
where $x$ satisfies the equation%
\begin{eqnarray}
4x^{3}+4i\left( E^{\prime }+1\right) x^{2}+(4c^{2}-4E^{\prime }-1)x+  \notag
\\
i(4c^{2}d^{2}E^{\prime }-2c^{2}d^{2}+2c^{2}-E^{\prime }) &=&0.
\end{eqnarray}%
The probability of the system being evolved from the initial state $|\Psi
_{0}\rangle $ to the state $|\Psi _{1}\rangle $ is $P(t)=|c_{1}\left(
t\right) |^{2}$. It depends on the evolution time $t$, the coupling
coefficient $c$, the overlap of the state $A|0\rangle ^{\otimes n}$ with the
ground state of the system, $d=\langle \varphi _{1}|A|0\rangle ^{\otimes n}$%
, and the eigenvalue $E^{\prime }+\frac{1}{2}$ of the state $|\Psi
_{2}\rangle $ and therefore can be expressed as $P(c,d,E^{\prime },t)$. It
reaches its maximal value as the runtime $t\sim \frac{1}{cd}$. The runtime
of the algorithm can be reduced if one can construct an unitary operator $A$
such that $A|0\rangle ^{\otimes n}$ is close to the ground state $|\varphi
_{1}\rangle $ of the system. Operator $A$ can be constructed in some simple
way to satisfy this condition in practice, e.g., in an application of our
algorithm for obtaining eigenstates and energy spectrum of water molecule
through nuclear magnetic resonance, we set $A=H_{d}^{\otimes n}$, where $%
H_{d}$ is the Hadamard matrices. The construction of operator $A$ can be
achieved using some state preparation techniques~\cite{wan, shende, mott,
berg}.

The coupling coefficient $c$ is related to the parameter $d$, here we set $%
c=d^{\alpha }$. In the following, we suppose the ground state energy of the
system $E_{1}=1$. By setting $\omega =1$ and $\varepsilon _{0}=0$, we study
the variation of the success probability of the algorithm $P(E^{\prime
},d,\alpha, t)$ with respect to the parameters $E^{\prime }$, $d$, $\alpha $
and $t$.

In Fig.~$2$, we set $d=0.01$ and plot the variation of $P$ with respect to $%
t $ and $E^{\prime }$ by setting $\alpha =1$ in Fig.~$2(a)$ and $\alpha =0$
in Fig.~$2(b)$, respectively. From the figures we can see that as $E^{\prime
}$ increases, $P$ becomes a periodic function with respect to the evolution
time $t$. And $P$ reaches unity quickly at small $E^{\prime }$ in the case $%
\alpha =1$ while at large $E^{\prime }$ in the case $\alpha =0$.

In Fig.~$3$, by setting $d=0.01$, we show the variation of $P$ versus $%
E^{\prime }$ at $t=\frac{\pi }{2}\frac{1}{cd}=\frac{\pi }{2}\frac{1}{%
d^{(1+\alpha )}}$ for $\alpha =0$, $0.5$, and $1$, respectively. From the
figure we can see that as the exponent $\alpha $ increases, $P$ reaches
unity quickly, and only at large $E^{\prime }$, the success probability $P$
can be close to unity for small exponent $\alpha $.

In Fig.~$4$, by setting $d=0.01$ and $E^{\prime }=5$, we show the variation
of $P$ versus the evolution time $t$ for $\alpha =0$, $0.5$, and $1$,
respectively. We can see that $P$ increases as $\alpha $ increases, $P$ is a
periodic function of $t$ and the period decreases as $\alpha $ increases.
And $P$ can be finitely large even in the case $\alpha =0$.

The runtime of the algorithm scales as $t\sim 1/\left( c|\langle \varphi
_{1}|A|0\rangle ^{\otimes n}|\right) $, we can make a guess on $t$ to run
the algorithm. And from Fig.~$4$, we can see that there is a large
probability for the success probability of the algorithm $P$ to be finitely
large with a guessed runtime $t$.

It is important to study the scaling of the exponent $\alpha $ with respect
to $d$ since the runtime of the algorithm is determined by $1/d^{(1+\alpha
)} $. In Table~I, we show the results for the variation of the exponent $%
\alpha $ vs. $d$ while keeping $P=0.99$ as $E^{\prime }=20$. From the Table
we can see that as $d$ increases, the exponent $\alpha $ decreases even to
zero at $d=0.4$. This means that the runtime of the algorithm scales as $1/d$%
, while in PEA, the success probability of the algorithm scales as $d^{2}$,
which means the algorithm has to be executed for $1/d^{2}$ times to obtain
the eigenstates. There is a quadratic speedup of our algorithm over the PEA
in this case. If we lower the success probability to $P=0.94$, $\alpha $ can
decrease to zero even at $d=0.1$, and the evolution time is reduced to $15$.
\begin{table}[tbp]
\caption{Results for variation of the exponent $\protect\alpha $ vs. $d$
while keeping $P=0.99$ as $E^{\prime }=20$. The runtime $t$ of the algorithm
is shown and compared with $1/d^{2}$, the efficiency of the phase estimation
algorithm.}
\begin{center}
\begin{tabular}{cccccccccccccccccc}
\hline
$d$ &  & $0.01$ &  & $0.02$ &  & $0.05$ &  & $0.1$ &  & $0.2$ &  & $0.4$ &
&  &  &  &  \\ \hline
$\alpha $ &  & $0.7$ &  & $0.6$ &  & $0.5$ &  & $0.35$ &  & $0.2$ &  & $0$ &
&  &  &  &  \\ \hline
$t$ &  & $3925$ &  & $815$ &  & $140$ &  & $35$ &  & $11$ &  & $4$ &  &  &
&  &  \\ \hline
$1/d^{2}$ &  & $10000$ &  & $2500$ &  & $400$ &  & $100$ &  & $25$ &  & $7$
&  &  &  &  &  \\ \hline
\end{tabular}%
\end{center}
\end{table}

Fig.~$5$ shows the variation of the exponent $\alpha $ vs. $d$ for $%
E^{\prime }=2,5,10,20$, respectively, while keeping $P=0.9$. We can see that
as $E^{\prime }$ increases, $\alpha $ decreases quickly and even reaches
zero at large $d$. This indicates that the runtime of the algorithm can
scale as $1/d$ while keeping a very high success probability $P=0.9$.

The time evolution operator $U(t)=\exp \left( -iHt\right) $ of the algorithm
can be implemented efficiently through the Trotter formula~\cite{nc} on a
quantum computer:

$U(t)=\left[ e^{-i\left( \frac{1}{2}\omega \sigma_{z}+H_{R}\right) t/M}%
e^{-i\left( c\sigma _{x}\otimes B\right) t/M}\right]^{M}+O\left( \frac{1}{M}\right) $, 
where $M$ is a large number.

\emph{Obtaining the energy spectrum of the system.}--In this algorithm, when
the transition frequency between the reference state and an eigenstate of
the system matches the frequency of the probe qubit, it contributes the most
to the decay of the probe qubit. By performing measurements on the probe
qubit in basis of $|0\rangle $ to obtain its decay probability, a peak in
the decay rate of the probe qubit will be observed. Therefore by varying the
frequency of the probe qubit or the eigenvalue of reference state $%
\varepsilon _{0}$, and run the algorithm, we can locate the transition
frequencies between the reference state and the eigenstates of the system~%
\cite{wan2}. Therefore obtain the energy spectrum of the system. The
detailed procedure is as follows.

First, we estimate the range of the energy spectrum of the system and set $%
\left[ \omega _{ini},\omega _{fin}\right] $ as the range of transition
frequency between the reference state and the eigenstates of the system. The
frequency range is then discretized into $q$ intervals, where each interval
has a width of $\Delta \omega =\left( \omega _{fin}-\omega _{ini}\right) /q$%
. The frequency points are set as $\omega _{k}=\omega _{ini}+k\Delta \omega $
$(k=0,\ldots ,q)$, and form a frequency set. We set the frequency of the
probe qubit to be $\omega _{k}$, and let the entire system evolve with the
Hamiltonian $H$ for time $t$. Then read out the state of the probe qubit by
performing a measurement on the probe qubit in basis $|0\rangle $. Repeat
the whole procedure many times to obtain the decay probability of the probe
qubit. Then set the probe qubit in another frequency and repeat the above
procedure until run over all the frequency points in the frequency set.

An application of our algorithm for obtaining eigenstates and the energy
spectrum of water molecule have been implemented experimentally through
nuclear magnetic resonance.

\emph{Discussion.}--For the Schr\"{o}dinger equation $H_{S}|\psi \rangle
=E|\psi \rangle $ of the system, an eigenvalue of $H_{S}$ can be obtained if
its corresponding eigenstate is known, and vice versa. Various methods based
on a guess state of the system have been developed to obtain the eigenstates
of the system including PEA. Here, we proposed a quantum algorithm for
obtaining the eigenstates of a system while the corresponding eigenvalue is
known. The algorithm is based on a physical phenomena that when a probe
qubit is coupled to a quantum system, the transition in the system that
resonates with the qubit contributes the most to the dynamics of the probe
qubit.

In the PEA for obtaining eigenstates and eigenenergies of a system, the
success probability of obtaining the $k$-th eigenstate $|\varphi _{k}\rangle
$ of the system is $|d_{k}|^{2}$, where $d_{k}$ is the overlap of the guess
state with $|\varphi _{k}\rangle $. And the number of times the algorithm
has to be run to obtain the eigenstate $|\varphi _{k}\rangle $ and its
corresponding eigenenergy is proportional to $1/|d_{k}|^{2}$. In our
algorithm, the overlap of the state $A|0\rangle ^{\otimes n}$ with the
eigenstate of the $i$-th energy level of the system is $d_{i}=\langle
\varphi _{i}|A|0\rangle ^{\otimes n}$. The runtime of the algorithm $%
1/|d_{i}|\leq t\leq 1/|d_{i}|^{2}$, the lower limit of the runtime of the
algorithm is consistent with the quantum speed limit for a system moving
from an initial state to a final state~\cite{gio}.

In this algorithm, all the eigenstates of the system are \textquotedblleft
labeled\textquotedblright\ by their eigenenergies, and an eigenstate of
interest is obtained by searching its \textquotedblleft
label\textquotedblright\ through inducing resonance with the probe qubit.
The probability of the system being evolved to the target state is amplified
by introducing a resonance between the probe qubit and a transition between
the reference state and the target state of the system. This is equivalent
to applying a quantum transformation on the system to achieve a quantum
speedup in searching the target state. In general, it can be viewed as an
amplitude amplification technique~\cite{grover2}. This explains why the
lower bound of the runtime of the algorithm is $1/|d_{i}|$, which is the
efficiency of the Grover's algorithm for the search problem~\cite{grover}.
And because of this property of the algorithm, for a given eigenvalue of the
system, all the corresponding eigenstates can be obtained, even for
degenerate eigenststes, in which case the adiabatic quantum evolution
algorithm cannot prepare all the eigenstates.

\begin{acknowledgements}
We thank L.-A. Wu, B. Sanders and D. Berry for helpful discussions. This work was supported by \textquotedblleft the Fundamental Research Funds for the Central Universities\textquotedblright\ of China and the National Nature Science Foundation of China~(Grants No.~11275145 and No.~11305120).
\end{acknowledgements}

\begin{figure}[tbp]
\includegraphics[width=0.8\columnwidth, clip]{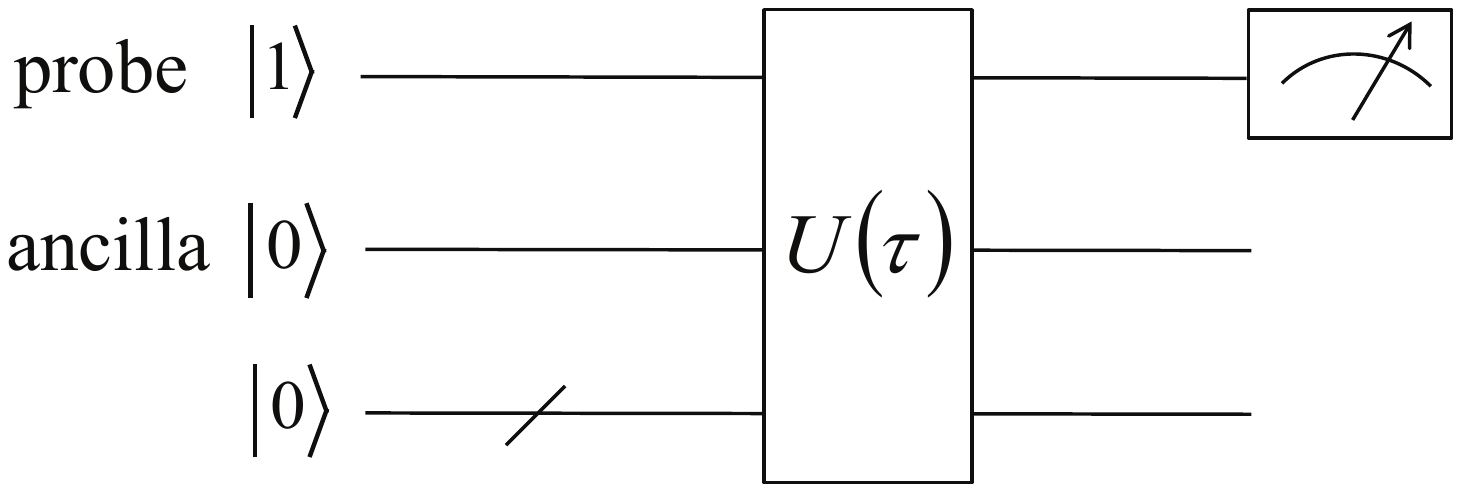}
\caption{Quantum circuit for obtaining the eigenstates of a physical system.
$U(\protect\tau )$ is a time evolution operator driven by a Hamiltonian
given in Eq.~($1$). The first line represents a probe qubit, the second line
is an ancilla qubit and the last $n$ qubits represent the quantum system.}
\end{figure}

\begin{figure}[tbp]
\begin{minipage}{0.48\linewidth}
  \centerline{\includegraphics[width=0.98\columnwidth, clip]{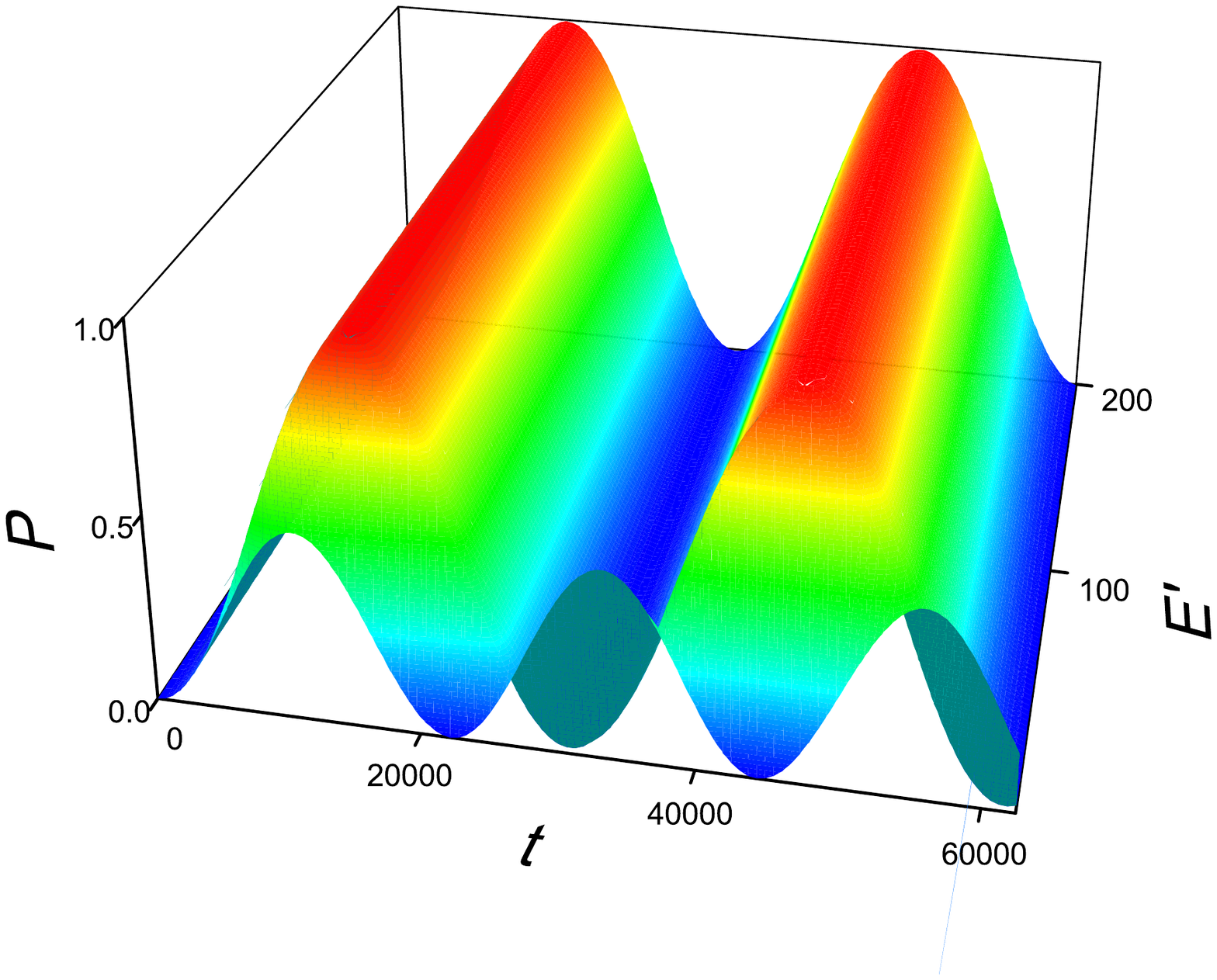}}
  \centerline{(a)}
\end{minipage}
\hfill
\begin{minipage}{0.48\linewidth}
  \centerline{\includegraphics[width=0.98\columnwidth, clip]{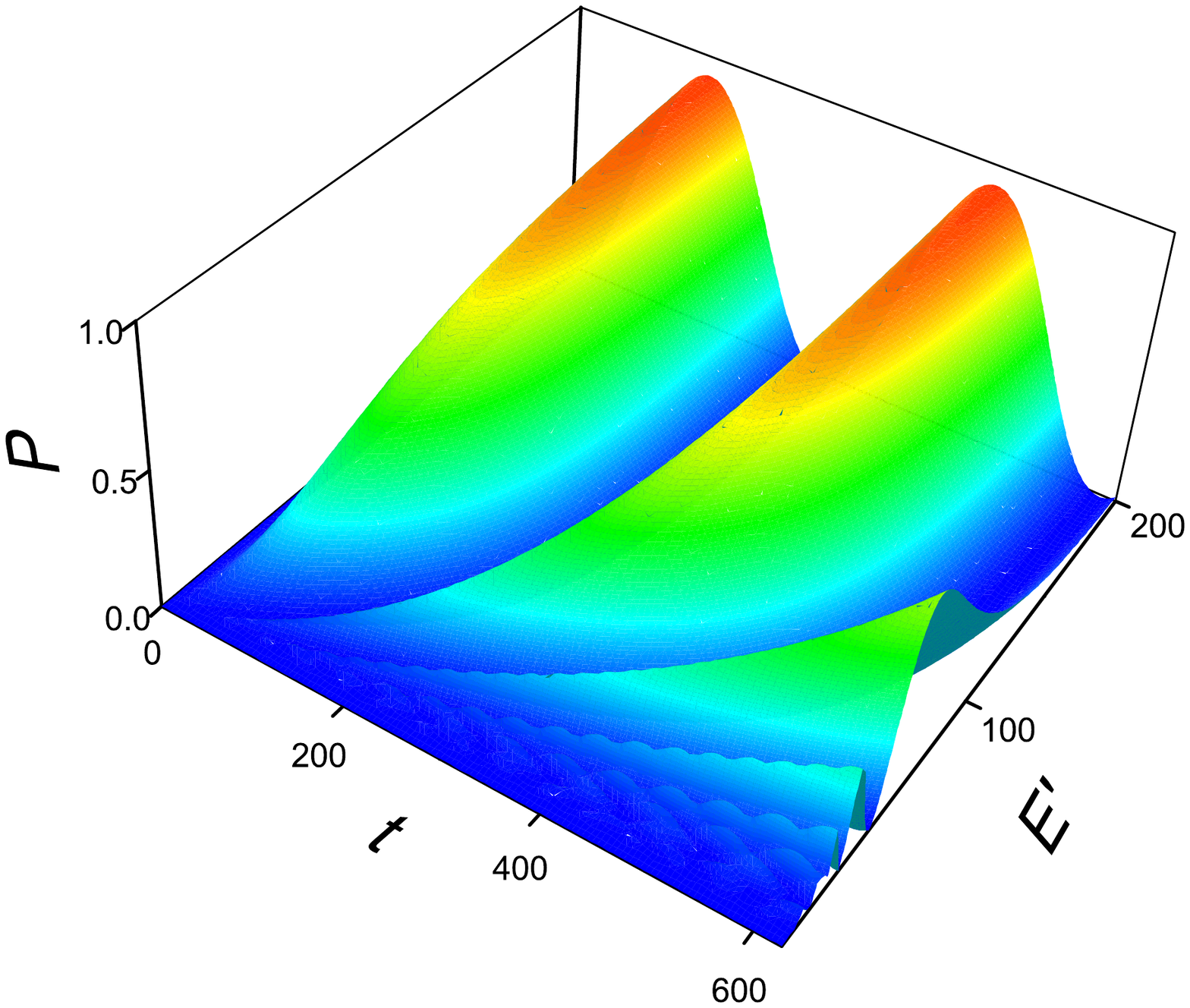}}
  \centerline{(b)}
\end{minipage}
\caption{(Color online)~The probability of the entire $(n+2)$-qubit system
being evolved from the state $|\Psi _{0}\rangle $ to the state $|\Psi
_{1}\rangle $ vs. the evolution time $t$ and $E^{\prime }$. In $(a)$ the
parameter $\protect\alpha =1$; and in $(b)$ $\protect\alpha =0$.}
\end{figure}

\begin{figure}[tbp]
\includegraphics[width=0.8\columnwidth, clip]{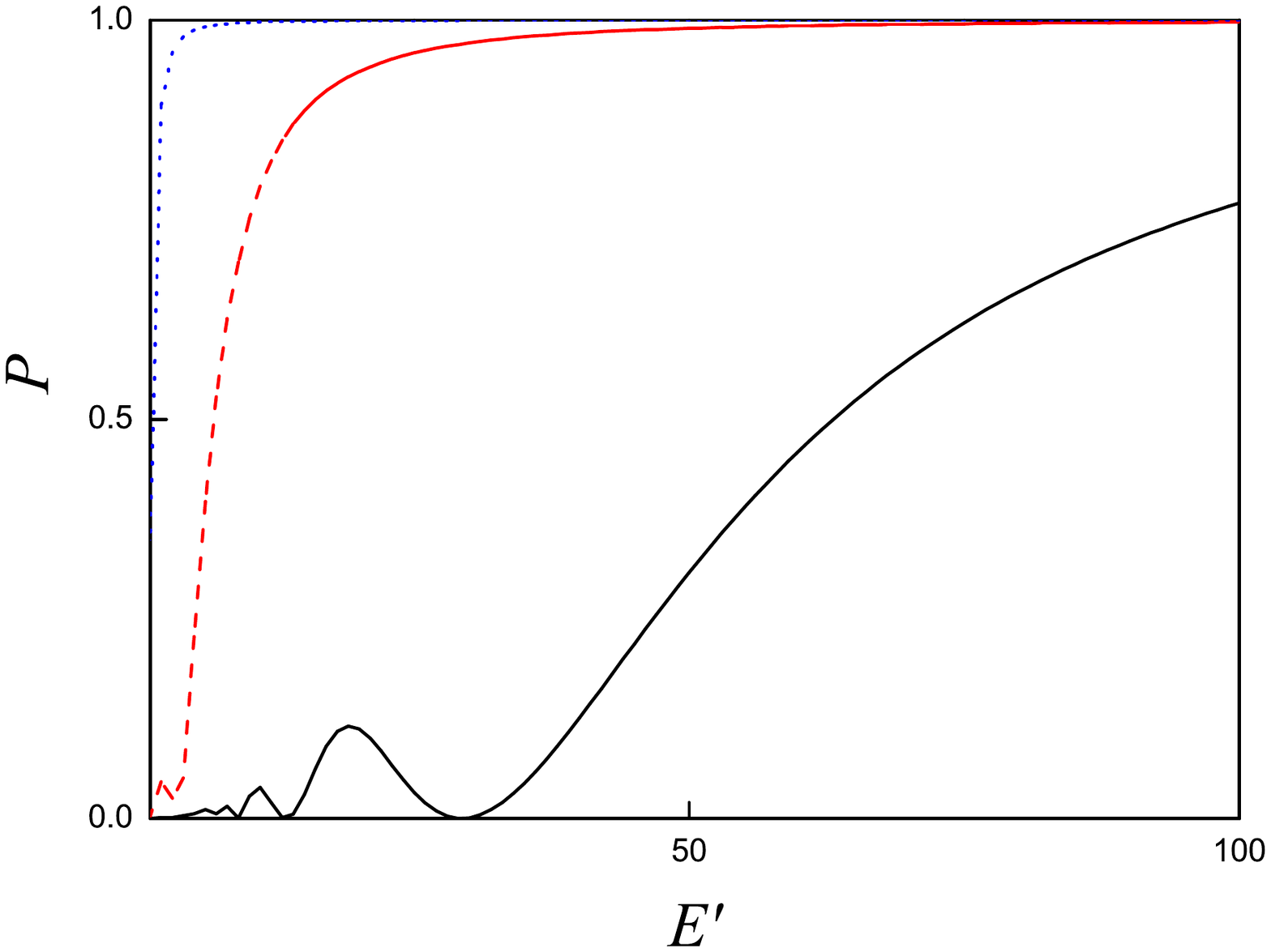}
\caption{(Color online)~The variation of $P$ vs. $E^{\prime }$ at $t=\frac{%
\protect\pi }{2}\frac{1}{d^{1+\protect\alpha }}$, for $\protect\alpha %
=0,0.5,1$ respectively by setting $d=0.01$. The black solid line shows the
results for $\protect\alpha =0$; the red dashed line shows the results for $%
\protect\alpha =0.5$; and the blue dotted line shows the results for $%
\protect\alpha =1$, respectively.}
\end{figure}

\begin{figure}[tbp]
\includegraphics[width=0.8\columnwidth, clip]{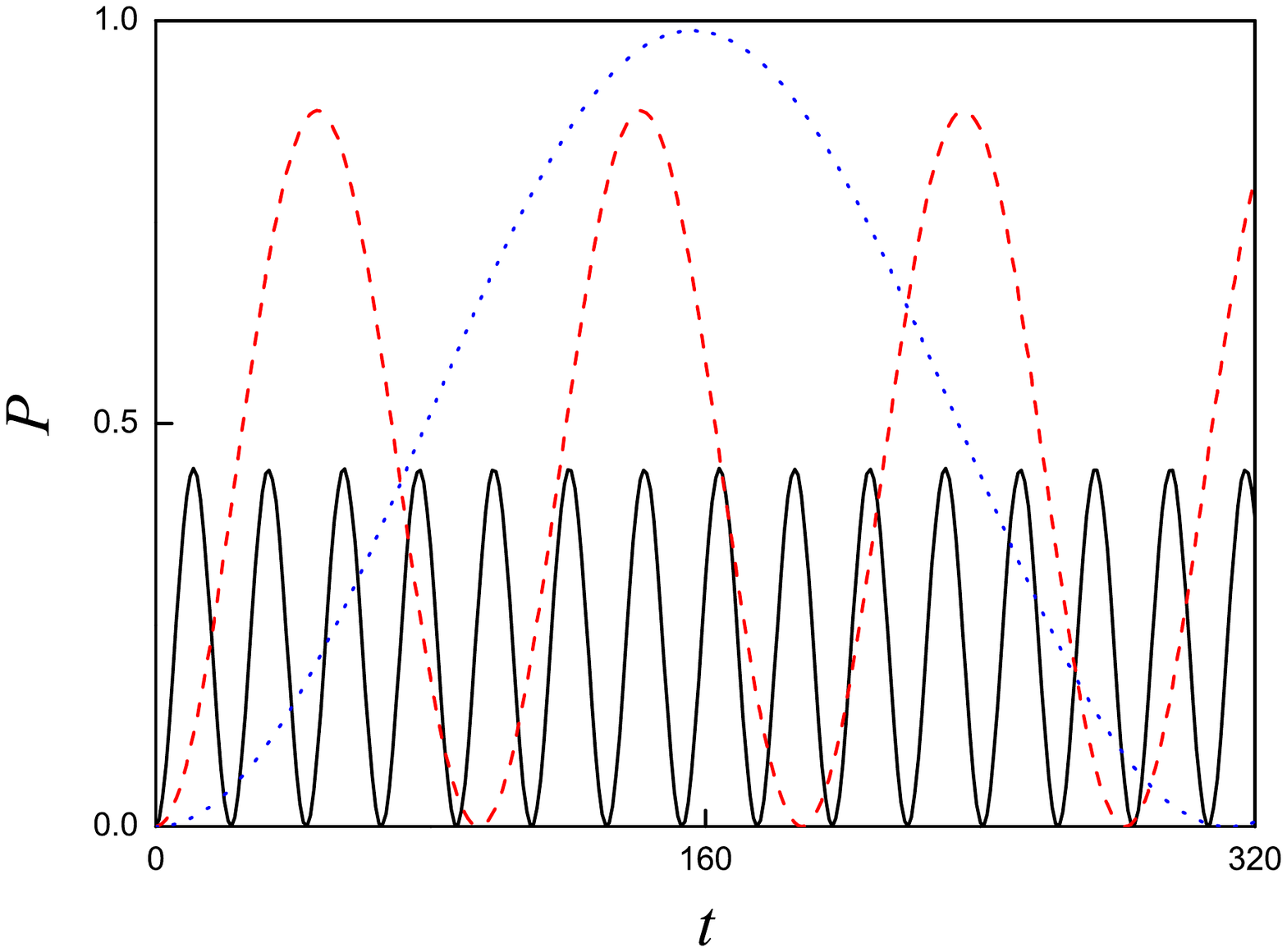}
\caption{(Color online)~The variation of $P$ vs. the evolution time $t$ for $%
\protect\alpha =0,0.5,1$ respectively, by setting $d=0.1$ and $E^{\prime }=5$%
. The black solid line shows the results for $\protect\alpha =0$; the red
dashed line shows the results for $\protect\alpha =0.5$; and the blue dotted
line shows the results for $\protect\alpha =1$, respectively.}
\end{figure}

\begin{figure}[tbp]
\includegraphics[width=0.8\columnwidth, clip]{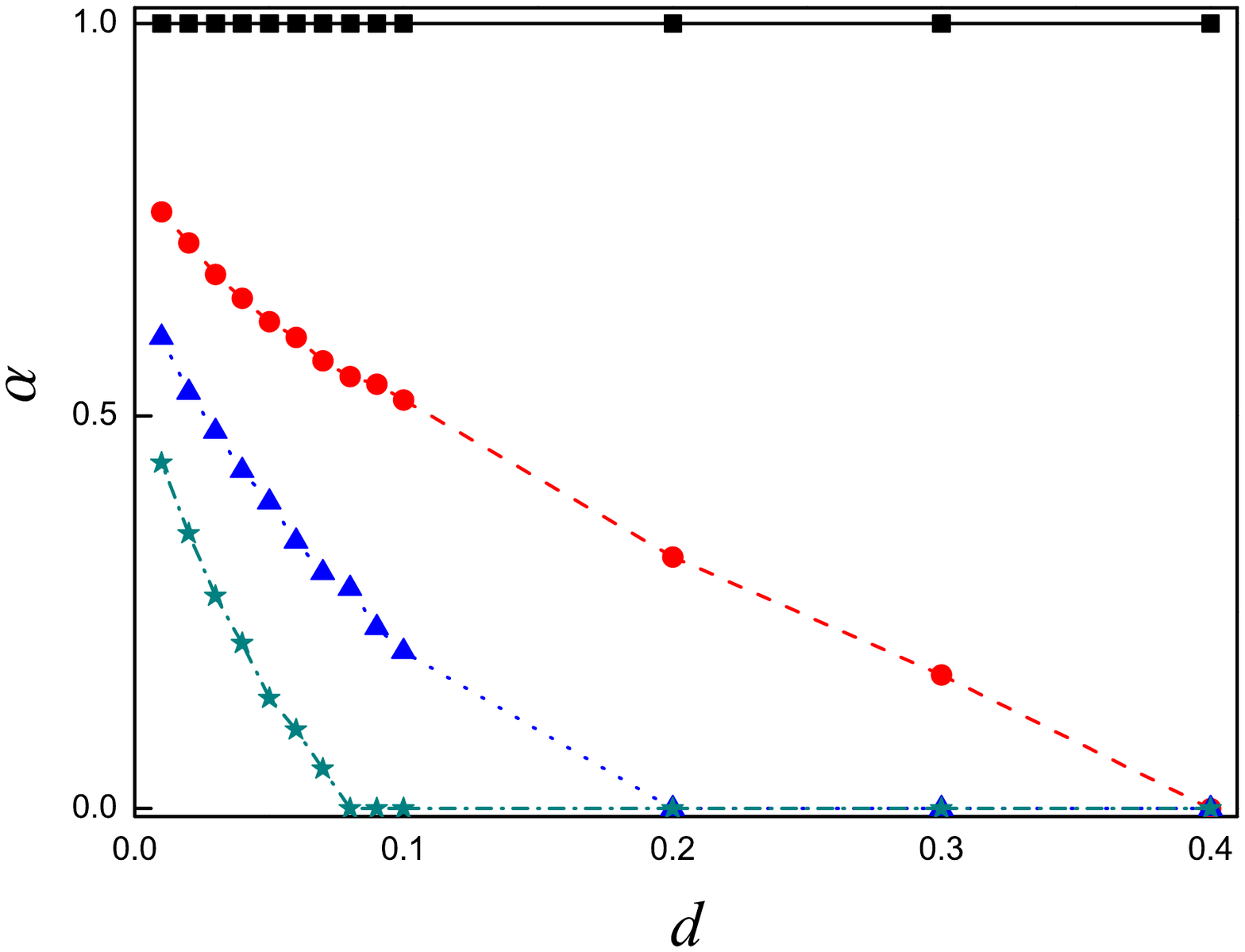}
\caption{(Color online)~The variation of the exponent $\protect\alpha $ vs. $%
d$ for $E^{\prime }=2,5,10,20$, respectively, while keeping $P=0.9$. The
black filled square shows the results for $E^{\prime }=2$; the red filled
circle shows the results for $E^{\prime }=5$; the blue filled triangle shows
the results for $E^{\prime }=10$; and the cyan filled star shows the results
for $E^{\prime }=20$, respectively.}
\end{figure}

\end{document}